\documentclass[twocolumn,english,natbib]{sigplanconf}
\usepackage{graphicx}
\usepackage{babel}
\usepackage{hyperref}
\begin{document}

\title{Density Adaptive Parallel Clustering}

\subtitle{A SS+tree based deterministic approach}

\authorinfo{Marcello La Rocca}{Scuola Superiore Sant'Anna}{marcellolarocca@gmail.com}
\maketitle
\begin{abstract}
In this paper we are going to introduce a new nearest neighbours based
approach to clustering, and compare it with previous solutions; the
resulting algorithm, which takes inspiration from both \textit{DBscan} and
minimum spanning tree approaches, is deterministic but proves simpler,
faster and doesn't require to set in advance a value for $k$, the
number of clusters.
\end{abstract}



\section{Introduction}

Unsupervised learning is the branch of machine learning that deals
with finding hidden structure in data; while supervised learning uses
previous information about a labeling of a training set to estimate
the labeling value for new input, data in unsupervised learning have
yet to be labeled or {}``classified''.

Approaches to unsupervised learning include: 
\begin{itemize}
\item clustering, also referred to as cluster analysis;
\item blind signal separation using feature extraction techniques for dimensionality
reduction (e.g., principal component analysis, independent component
analysis, non-negative matrix factorization, singular value decomposition).
\cite{acharyya2008new}
\item Neural network models, like self-organizing map (SOM) and adaptive
resonance theory (ART).
\end{itemize}
In clustering, the goal is to divide data points into homogeneous
subsets, called clusters, such that objects in the same subset or
\textbf{cluster }are more similar (according to a specific measure)
to each other rather than to those in other subsets.

Many clustering algorithms have been proposed during the years, some
more efficient than others, but it is difficult to define an objective
definition of what a good cluster should be, and the decision about
which of these algorithm to use deeply depend on the characteristics
of the data set to be partioned.

\begin{figure}[tbh]
\includegraphics[width=85mm]{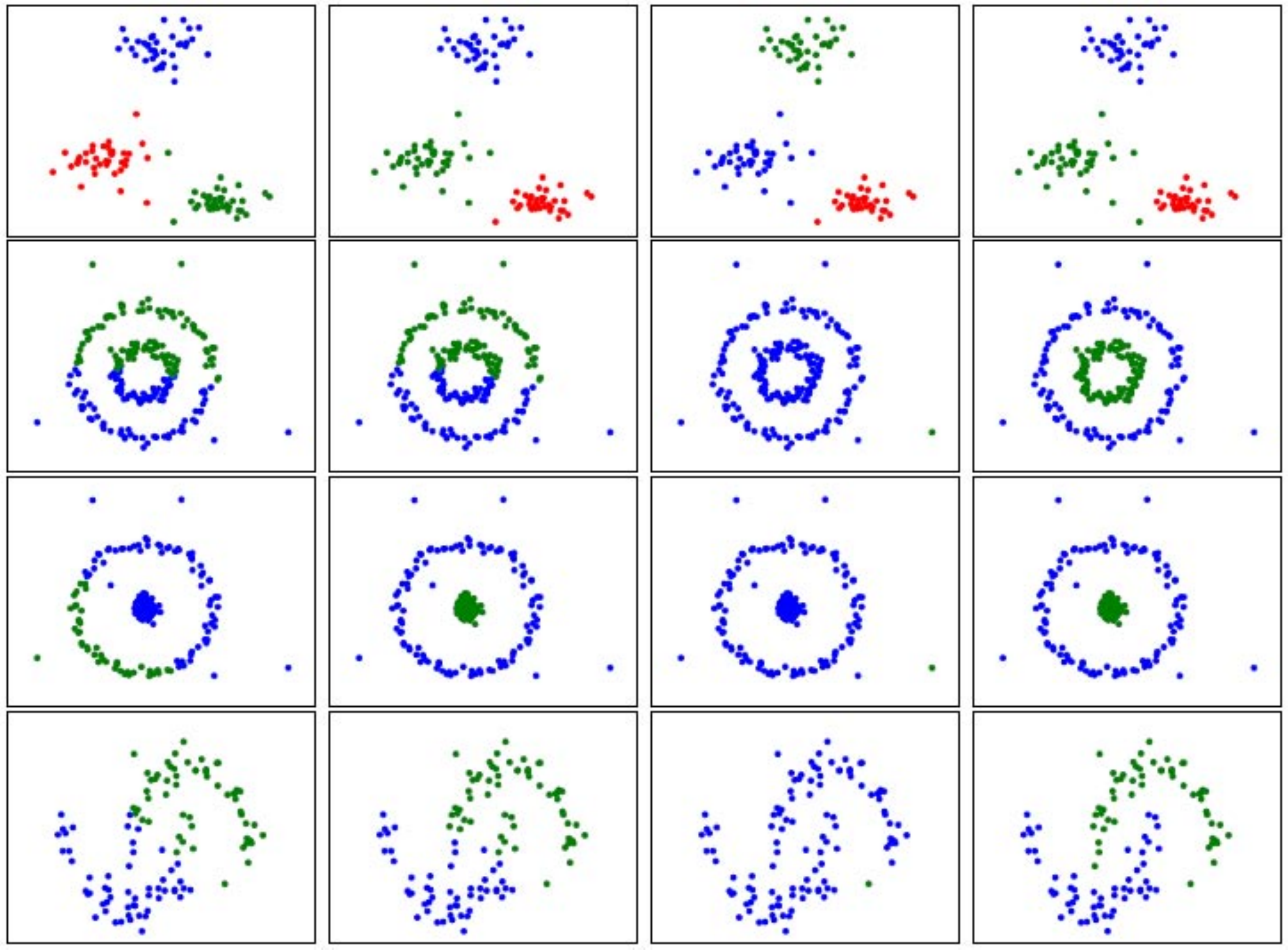}

\caption{Comparison of k-Means (left), MeanNN (center left), single link (center
right) and ITM (right) on four synthetic datasets, as shown in \cite{muller2012information}}

\end{figure}

\section{Related Work}

Most clustering algorithms can be categorized based on their cluster
model, although not every clustering algorithm provide a model for
their clusters; some of the most important categories are:
\begin{itemize}
\item \textbf{Centroid-based clustering}: clusters are represented by a
central point, which may not necessarily be a member of the data set.
Since this optimization problem is known to be NP-hard, the common
approach is to search only for approximate solutions. The most commonly
used centroid-based clustering algorithm is Lloyd's algorithm, usually
referred to as the \textit{k-Means} algorithm \cite{lloyd1982least,macqueen1967some};
\textit{k-Means} often works well in practice, but its main limitation
is the restriction in cluster shape, that is always forced to be convex.
One method that overcomes this limitation is spectral clustering \cite{ng2002spectral,shi2000normalized},
which solves a graph partitioning problem on a similarity graph based
on the data, and hence its performance is heavily affected by the
particular choice of graph construction and similarity measure. 
\item \textbf{Hierarchical, or connectivity based, clustering}: based on
the core idea of objects being more related to nearby objects than
to objects farther away. These algorithms connect \char`\"{}objects\char`\"{}
to form \char`\"{}clusters\char`\"{} based on their distance. A cluster
can be described largely by the maximum distance needed to connect
parts of the cluster. The main downside for these algorithms is their
asynthotic running time, which for the simplest methods results $O\left(n^{3}\right)$;
more sophisticated methods, like \textit{SLINK} \cite{sibson1973slink}
and \textit{CLINK} \cite{defays1977efficient}, have a worst case
running of $O\left(n^{2}\right)$.

\begin{itemize}
\item \textbf{MST-based clustering}: it's a subset of hirarchical clustering,
and it's been an important filed of research for decades, since 1969
Gower and Ross paper on single-link agglomerative clustering \cite{gower1969minimum},
a minimum spanning tree-based approach in which the largest edge is
removed until the desired number of components is reached. This criterion
is further studied by Zahn \cite{zahn1971graph}: in his work, only
edges that are longer than other edges in the vicinity are cut, although
this approach requires tuning several constants by hand. More recently,
Grygorash et al. \cite{grygorash2006minimum} proposed a hierarchical
MST-based clustering approach that iteratively cuts edges, merges
points in the resulting components, and rebuilds the spanning tree,
while in 2012 Muller et al. \cite{muller2012information} proposed
an efficient algorithm that infers cluster memberships by direct optimization
of a non-parametric mutual information estimate between data distribution
and cluster assignment.
\end{itemize}
\item \textbf{Distribution-based clustering}: Distributional clustering,
works only under the assumption that the distribution of the data
is known explicitly (for example as word counts), which is often not
the case in practical situation. Later, in 2005, Banerjee et al. \cite{banerjee2005clustering}
introduced the concept of Bregman Information, generalizing mutual
information of distributions, and showed how this leads to a natural
formulation of several clustering algorithms. Barber \cite{barber2005kernelized}
constructs a soft clustering by using a parametric model of conditional
probability, and the framework of mutual information based clustering
was recently extended to non-parametric entropy estimates by Faivishevsky
and Goldberger \cite{faivishevsky2010nonparametric}. They use a nearest
neighbor based estimator of the mutual information, called MeanNN,
that takes into account all possible neighborhoods, therefore combining
global and local influences. 
\item \textbf{Density-based clustering}: clusters are defined as areas of
higher density than the remainder of the data set. Objects in these
sparse areas - that are required to separate clusters - are usually
considered to be noise and border points. The most popular density
based clustering method is \textit{DBSCAN} \cite{ester1996density},
which (similarly to linkage based clustering) works by connecting
points within certain distance thresholds, the only difference being
that it only connects points that satisfy a density criterion. An
interesting property of \textit{DBSCAN} is that its complexity is
fairly low - it requires a linear number of range queries on the database
- and it is deterministic in nature (except for border points), hence
it doesn't have to be run multiple times. \textit{OPTICS} \cite{ankerst1999optics}
is a generalization of \textit{DBSCAN} that removes the need to choose
an appropriate value for the range parameter, and produces a hierarchical
result related to that of linkage clustering. \textit{DeLi-Clu} \cite{achtert2006deli},
\textit{Density-Link-Clustering}, combines ideas from single-linkage
clustering and \textit{OPTICS}, eliminating the parameter entirely
and offering performance improvements over \textit{OPTICS} by using
an R-tree index for nearest neighbours search. The key drawback of
\textit{DBSCAN} and \textit{OPTI}CS is that they expect some kind
of density drop to detect cluster borders. Moreover they can not detect
intrinsic cluster structures which are prevalent in the majority of
real life data. A variation of \textit{DBSCAN}, \textit{EnDBSCAN}\cite{roy2005approach}
efficiently detects such kinds of structures. 
\end{itemize}
An increasingly crucial issue is performance: the rise of big data
led to a need to process larger and larger data sets, which in turn
led to an increasingly tendency to trade semantic meaning of the generated
clusters for performance. To cope with this need, pre-clustering methods
such as canopy clustering \cite{mccallum2000efficient} were introduced
to process huge data sets efficiently, even if the resulting \char`\"{}clusters\char`\"{}
are merely a rough pre-partitioning of the data set, and those partitions
need to be analyzed in a later step with the existing slower methods
above.

\section{A new approach, inspired by the past}

Our solution has been inspired by both DBScan and MST approaches;
while the first one is very fast and well performing, it requires
a bit of tuning for each dataset because two parameters need to be
set, a threshold and the radius of the scanning area; MST based approaches,
on the other hand, achieves better results with ill-shaped clusters
{[}see figure 1{]}\cite{muller2012information}, but it is so slow
to be unpractical for moderately large base of data (more than 10K
elements).

Noticing the flaws of the MST solution, we decided to try to gain
the same benefits using a different way to improve performance. The
key observation is that, if we are going to apply Kruskal algorithm
to build the MST of an {}``euclidean'' graph, i.e. a fully connected
graph of points in the plane, where edge weight coincides with Euclidean
distance, at each step the shortest path between a vertex in the cut
and a vertex outside the cut, is the edge connecting the two of them,
and the shortest cut-edge is the one among cut and non-cut vertices
with the smallest distance. 

\begin{figure}[tbh]
\includegraphics[width=85mm]{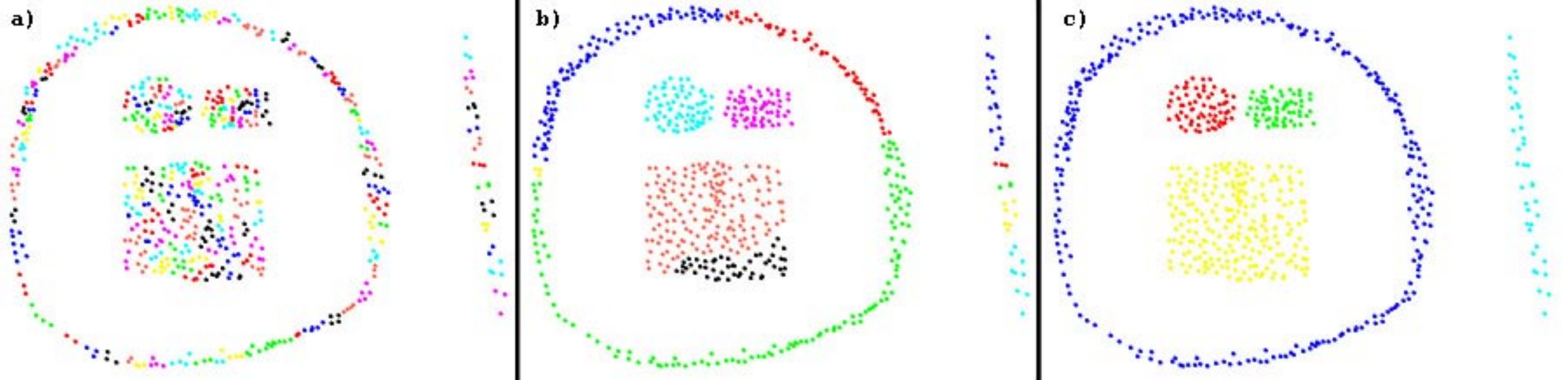}

\caption{Clusting resulting from the naive nearest neighbour approach, with
differents value for parameter $m$ (respectively $m=1$, $m=2$,
$m=3$) }
\end{figure}

With this in mind, the first, naive, approach would obviously be to
scan every point in the dataset and merge its cluster with it's nearest
neighbour's one; to perform this operation quickly we decide to use
\textit{SS+Trees} \cite{kurniawati1997ss+} for nearest neighbours
computation, and weighted UnionMerge with path compression to keep
track of the elements' clusters (see an analysis of the running time
below). However, the result was not particularly encouraging, and
it is not difficult to see why: most points are each other nearest
neighbours, so we end up with a high fragmentation, due to a lot of
very small clusters as you can see in the example from figure {[}Figure
2a{]}.

By incrementing $m$, the number of nearest neighbours considered
for each data point, the results improves dramatically, as you can
see in {[}Figure 2b{]} ($m=2$) and {[}Figure 2c{]} ($m=3$): the
clustering of this particular datasets looks just right, and this
version of algorithm already succeeds even where \textit{k-means}
and \textit{NN-mean} are forced to failure by their own nature as
shown in {[}Figure 3{]}.

However, even this improved version has a flaw. Consider the dataset
in {[}Figure 4{]}: a single point $z$ has been strategically added
to the dataset in {[}Figure 3{]}, so that it is approximately at the
same distance from blue and red clusters; this in turn means that
the two closest points to the newly added one, denoted as $x_{z}^{r}$
and $x_{z}^{g}$, lie one in the red cluster, and one in the green
one. Therefore, when $z$ is examined, first its cluster $c_{z}$,
which at that point only contains $z$ itself, will be merged with
$x_{z}^{r}$'s one, creating a new cluster $\hat{c_{z}}$, and then
$\hat{c_{z}}$ will be merged with $x_{z}^{g}$'s cluster, so that
$x_{z}^{r}$ and $x_{z}^{g}$ will land in the same cluster, as shown
in {[}Figure 4a{]}. This problem is not new for clustering algorithms:
it is commonly known as the \textit{single-link effect}, and will
affect every dataset with dense clusters and some sparse points, especially
if they are approximatively equidistant from 2 dense clusters. As
a generalization, this unwanted merge of what should be separate clusters
will be caused by chains of $j$ sparse points liyng at the same approximate
distance from 2 dense clusters, when the smoothing parameter $s\geq j+1$
{[}Figure 4b{]}.

There is no way to solve this problem by tuning the smoothing parameter,
so we needed an alternative approach.

\begin{figure}[tbh]
\includegraphics[width=85mm]{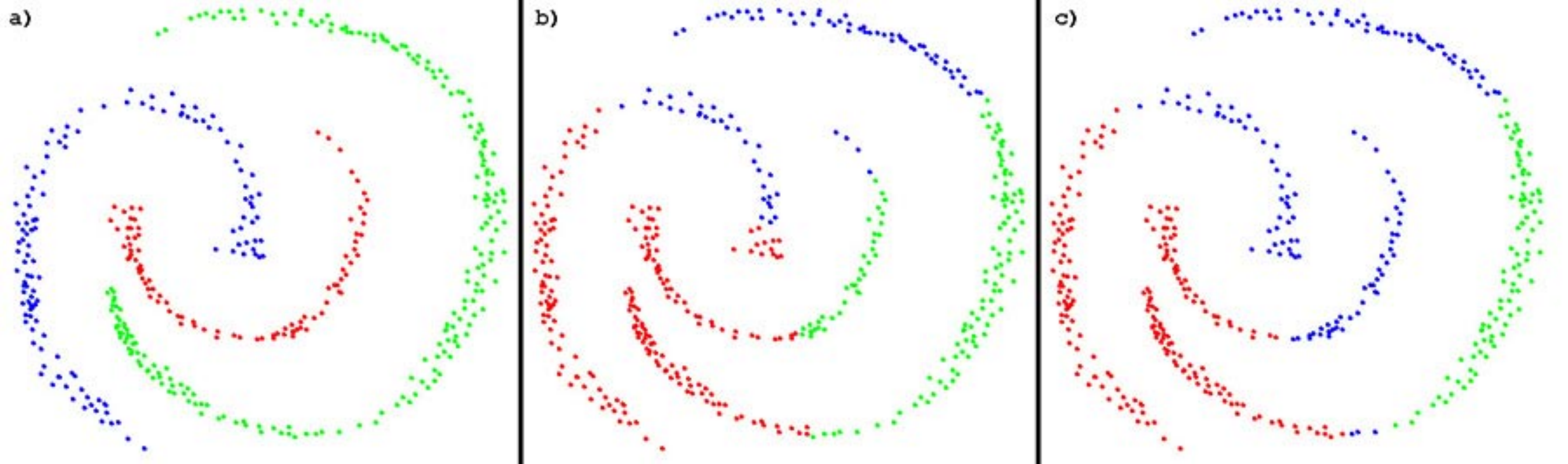}

\caption{Comparison between 3-nearest neighbours algorithm (a) and kmeans (b)
and meann NN (c) algorithms with $n\_cluster=3$}

\end{figure}

\section{An improved version}

The previous algorithm performs very well on most datasets, with a
little bit of tuning. However, tuning becomes incresingly complicated
as the number of points in the dataset grows, and it perniciously
shows the tendency to merge together well separated clusters, if a
few noise points lies between them. Nonetheless, it represents a fast
alternative when noise in the data set is very low.

We knew, however, that - at the cost of some performance - there existed
a better solution.

\begin{figure}[tbh]
\includegraphics[width=85mm]{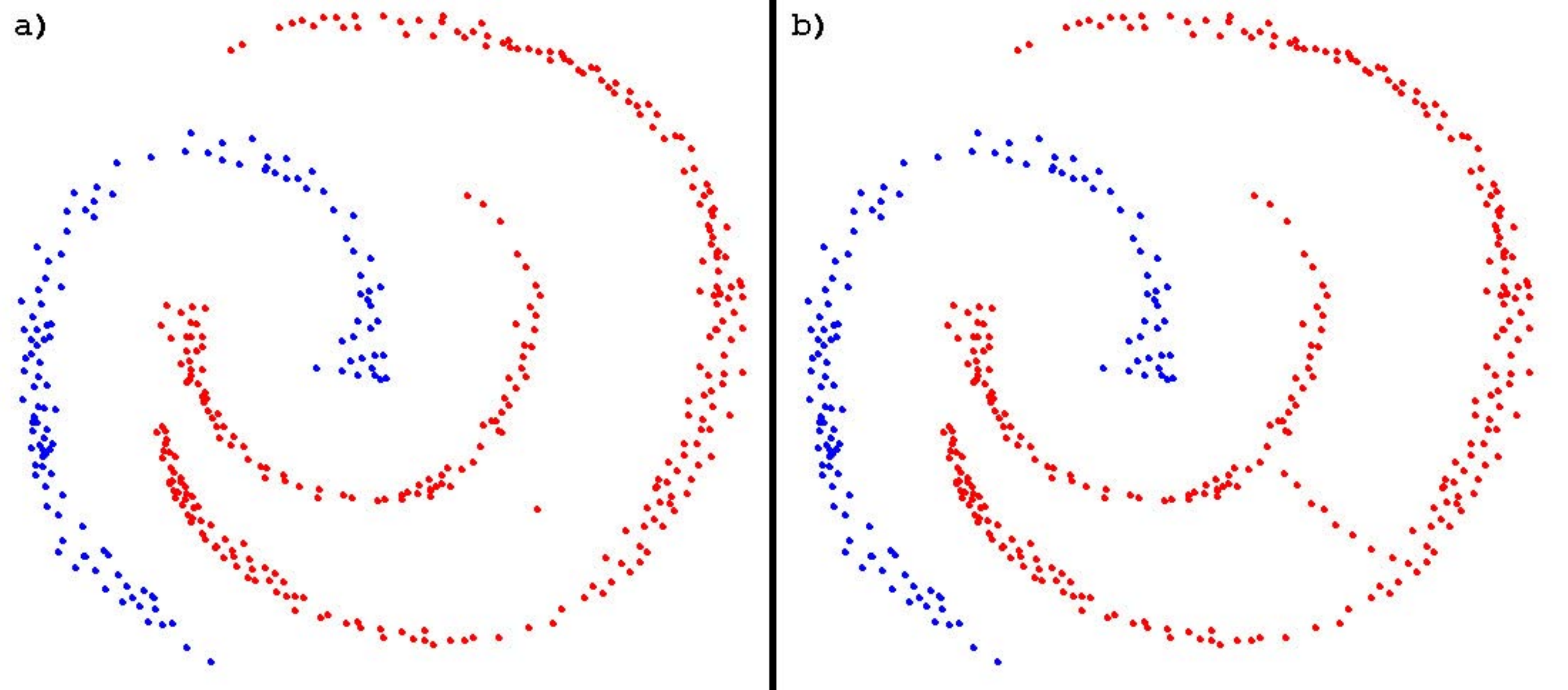}

\caption{Single link effect on the naive clustering algorithm}

\end{figure}

What the naive algorihtm was lacking is a way to discriminate noise
or, to put it another way, to rule out points too far from our potential
clusters.

There were many ways to try to do so - for example running a noise
detection algorithm as a first step, and ignore those points identified
as noise in our main algorithm.

Instead, we decided to take inspiration from \textit{DBscan} \cite{ester1996density},
a breaktrough in clustering algorithms. \textit{DBscan} considers
two points $p$ and $q$ to be part of the same cluster if they aren't
further apart than a certain distance $\varepsilon$ and if either
$p$ or $q$ is surrounded by at least $m$ points; such points are
called \textit{density-reachable}. The idea behind \textit{DBscan}
is the following: for each point $p$, scan an area of radius $\varepsilon$
around $p$, and if at least $m$ points are found in that area, all
those points are \textit{density-reachable} from p, and so they belong
to the same cluster. This also implies that if there are two points
$p,\, w$ such that there is a sequence $p=p_{1},\: p_{2},\,...\,,\: p_{n}=w$
where $p_{i+1}$ is \textit{density-reachable} from $p_{i},\:\forall i=1...\: n-1$,
then $p$ and $w$ are in the same cluster; such points are called
\textit{density-connected}, and all \textit{density-connected} points
belong to the same cluster.

While \textit{DBscan} shows several advantages in comparison to \textit{k-means}
(it can find clusters of any shape, it deals with noise and the number
of clusters doesn't need to be set \textit{a priori}), there are three
main problems with \textit{DBscan}: 
\begin{enumerate}
\item Both $\varepsilon$ and $m$ are input parameters for \textit{DBscan},
so they must be decided by the user in advance, and a fair amount
of tuning might be needed to get both of them right;
\item It doesn't scale well to higher dimentional data; {[}besides it requires
either $O\left(n^{2}\right)$ time or space{]}
\item If the dataset has areas with very different density, then it works
very poorly on at least one of them, since it is impossible to find
a good compromise for the values of $\varepsilon$ and $m$ that works
for all the densities.
\end{enumerate}
Our goal is to introduce the concept of \textit{density-reachablility}
in our naive algorithm, but in an adaptive way, such that the $\varepsilon$
and $m$ parameters are inferred independently for each single point
in the dataset, in order to cope with areas with great differences
in their density, while still being able to spot and discard noise.

We ended up with an approach similar to DeLi-Clu\cite{achtert2006deli},
with a two steps algorithm: first it scans the dataset, creating a
pre-clustering partition, and then for each subset in this partition,
estimate a suitable value for the radius $\varepsilon$ of the area
to scan to look for neighbours, and uses it in a \textit{DBscan} alike
step.

In particular, during the second phase, for each point with at least
$m$ neighbours within its scan area, all its neighbours inside the
radius of the area will end up in the same cluster as the point
itself. To efficiently perform these two operations, we decided to
use \textit{SS+trees} \cite{kurniawati1997ss+} for spatial queries,
and \textit{UnionFind} structures to keep track of the clusters: initially
every point it's a cluster of its own, then clusters keep being merged
as similarities between points are discovered.

These choices bring several advantages, especially as performance
is concerned. As a matter of fact, as we already suggested in the
introduction section, a critical problem with modern large dataset
is performance, due to the huge amount of data that has to be processed.

Using the \textit{UnionFind }approach allows a map reduce strategy:
we can divide the data space into (slightly overlapping) regions,
compute the clustering for each of these regions (map step) and then
merge the set resulting from the union find algorithm for each point
over the regions in which it appears (reduce step). Each region can
thus be assigned to a different processor, since no shared data structure
is needed, and reduction can be performed incrementally, as new nodes
\textit{pubblish} their results, so the computation can support a
high degree of parallelization. The additional advantage in such a
strategy is of course that we can use different parameters for each
of these regions, taking into account differences in density as no
algorithm could before, and also that \textit{UnionFind} and \textit{SS+trees}
can run and be computed on a smaller set of points: at the cost of
a little space overhead due to the overlapping nature of the regions,
the total performance sensibly improves as the number of regions grow.

The key point, of course, is how to find these regions in the first
place. If, from a computational point of view, a naive partitioning
in regular, same-sized regions could work just as fine, our final
goal is to devise the partitioning such that it helps us to cope with
possible difference in density. The best approach turned out to be
using pre-clustering, in particular step 1 consist of applying \textit{canopy
clustering} \cite{mccallum2000efficient} on the complete data set
with an approximate metric to fastly compute the partioning, then
compute the $\varepsilon$ parameter based on the points in each partition,
and finally build regions for step 2 around these partitions (tipically
the smallest box or sphere containing the partition, with at least
$m$ points).

As shown by Apache Mahout project (\url{https://cwiki.apache.org/confluence/display/MAHOUT/Canopy+Clustering}),
\textit{canopy clustering} support a high degree of parallelization
and map reduce strategies as well, since every node can run indipendently.

Another relevant problem is the dimentionality curse: again, \textit{SS+trees}
do scale better than \textit{R-trees} considering both wasted space
(reduced by using spheres instead of boxes) and efficiency (improved
by computing an approximation to the smallest enclosing sphere - whose
computation is exponential in the number of dimentions).

The final algorithm can be summarized in the following steps:
\begin{itemize}
\item \texttt{First step, run canopy clustering to compute }

\texttt{pre-clusters,dividing the problem around these } 

\texttt{partitions;}
\item \texttt{Second step {[}Map{]}, process each partition }

\texttt{separately;}
\item \texttt{Third step {[}Reduce{]}, merge the sets of the union find
algorithm for each vertex produced by step 2.}
\end{itemize}

\section{Performance}

There are four different sub-algorithms contributing to the total
running time of the clustering algorithm:
\begin{enumerate}
\item \textit{Canopy clustering} algorithm performance will be strongly
dependent on the choice of the metric; our goal is to keep this step
as close to $O\left(n\right)$ as possible;
\item The creation of a \textit{SS+Tree}, which for $k$ points require
$O\left(k\,\log k\right)$ time; it also require $O\left(k\right)$
extra space to retain the information;
\item For each point:

\begin{enumerate}
\item A query to find all the points within the scanning area is executed;
this search operation is certainly $\Omega\left(\log k\right)$, because
that's the lower bound for a nearest neighbour search on a \textit{SS+tree},
but it can also be as bad as $O\left(k\right)$ in the worst case,
because if the area is big enough it could include all the points
in the data set. Typically, however, the area radius is chosen so
that a small number of points ($m$ on average, where $m$ is the
smoothing parameter defined above) will be included in the scanning
area, so small it can considered constant on average, so the average
running time for each search operation is $O\left(\log k\right)$;
\item The \textit{UnionMerge} procedure will be executed for every one of
its neighbours within the scanning area, so, on average, it will be
called $m$ times; the average running time of $m$ \textit{UnionMerge}
operation with \textit{union rank} and \textit{path compression} is
$O\left(m\,\alpha\left(k\right)\right)$, where $\alpha\left(n\right)$
is the inverse \textit{Ackerman} function, that grows so slowly that
it can be considered lower than 5, and so constant, for any practical
value of $n$.
\item The \textit{UnionFind} data structure used in step (b) requires $O\left(k\right)$
time time to be built and $O\left(k\right)$ extra space;
\end{enumerate}
\item Therefore, the total running time of the second step is

 $O\left(mk\,\alpha\left(k\right)+k\,\log k\right)$
 
for $k$ points, which - if $m$ is constant - becomes 

$O\left(k\,\alpha\left(k\right)+k\,\log k\right)=O\left(k\,\log k\right)$.
\end{enumerate}
Without the map reduce parallelization strategy, the \textit{SS+tree}
and \textit{UnionFind} structure will be constructed once for the
whole dataset, so $k=n$ and the total running time will be 

$O\left(n+n\,\log n+n\,\alpha\left(n\right)+n\,\log n+n\right)=O\left(n\,\log n\right)$;

the extra space required in this case will be $O\left(n\right)$.

Considering the execution of step 2 in parallel on each region, if
there are $z$ regions with at most $w$ elements, the worst running
time for the biggest region will be $O\left(w\,\alpha\left(w\right)+w\,\log w\right)$,
and the running time to merge the union sets will be, in the worst
case, $O\left(zw\right)$, which in turn is $\Omega\left(n\right)$,
but depending on the overlapping factor of the regions can be as big
as $n^{2}$. If we assume that each region has at least $m$ points,
and we force each point to appear in at most $m$ regions, then we
can assume $O\left(zw\right)=O\left(n\right)$.

The running time becomes: 

$O\left(n+w\,\log w+w\,\alpha\left(w\right)+w\,\log w+w+zw\right)=O\left(n+w\,\log w\right)$;

the extra space needed will be proportional to $O\left(zw\right)$
ans so, under the assumption above, still $O\left(n\right)$.

\subsection{Comparison to other clustering algorithms}
In our experiments, our parallel algorithm proved to be as fast as \textit{k-Means}, twice as fast as \textit{DBscan}, and an order of magnitude faster than \textit{Mean NN}.
While \textit{k-Means}' performance is heavily influenced by the number of clusters sought, there is no noticeable difference in our algorithm's running time when parameters are tuned.
It is possibile to take a look at the code for the algorithms tested and our testing environment \url{https://bitbucket.org/mlarocca/clustering}, and a demo version is also available as a web-app here \url{http://nnn-clustering.appspot.com/static/index.html}.

\section{Conclusions and future work}

The results of the approach presented proved to be very robust, allowing
the clustering of huge data sets. Our goal is to further pursue this
road, trying to perfect final clustering with a method more sensitive
to cluster borders and density.

You can use either Bib\TeX{}:

\bibliographystyle{plainnat}
\addcontentsline{toc}{section}{\refname}\bibliography{dap-clustering}

\end{document}